# Anisotropic long-range spin transport in canted antiferromagnetic orthoferrite YFeO$_3$

Shubhankar Das[1], A. Ross[2,$], X. X. Ma[3,$], S. Becker[1], C. Schmitt[1], F. van Duijn[4,5], F. Fuhrmann[1], M.-A. Syskaki[1], U. Ebels[4], V. Baltz[4], A.-L. Barra[5], H. Y. Chen[3], G. Jakob[1,6], S. X. Cao[3,§], J. Sinova[1], O. Gomonay[1], R. Lebrun[2], and M. Kläui[1,6,7*]

[1]Institute of Physics, Johannes Gutenberg University Mainz, Staudingerweg 7, 55128 Mainz, Germany

[2]Unité Mixte de Physique CNRS, Thales, Université Paris-Saclay, Palaiseau 91767, France

[3]Department of Physics, Materials Genome Institute, International Center for Quantum and Molecular Structures, Shanghai University, Shanghai, 200444, China

[4]Univ. Grenoble Alpes, CNRS, CEA, Grenoble INP, SPINTEC, F-38000 Grenoble, France

[5]Laboratoire National des Champs Magnétiques Intenses, CNRS-UGA-UPS-INSA-EMFL, F-38042 Grenoble, France

[6]Graduate School of Excellence Materials Science in Mainz, Staudingerweg 9, 55128 Mainz, Germany

[7]Center for Quantum Spintronics, Norwegian University of Science and Technology, Trondheim 7491, Norway

*klaeui@uni-mainz.de

§sxcao@shu.edu.cn

$these authors are contributed equally

In antiferromagnets, the efficient propagation of spin-waves has until now only been observed in the insulating antiferromagnet hematite, where circularly (or a superposition of pairs of linearly) polarized spin-waves propagate over long distances. Here, we report long-distance spin-transport in the antiferromagnetic orthoferrite $YFeO_3$, where a different transport mechanism is enabled by the combined presence of the Dzyaloshinskii-Moriya interaction and externally applied fields. The magnon decay length is shown to exceed hundreds of nanometers, in line with resonance measurements that highlight the low magnetic damping. We observe a strong anisotropy in the magnon decay lengths that we can attribute to the role of the magnon group velocity in the propagation of spin-waves in antiferromagnets. This unique mode of transport identified in $YFeO_3$ opens up the possibility of a large and technologically relevant class of materials, i.e., canted antiferromagnets, for long-distance spin transport.

The field of antiferromagnetic spintronics seeks to functionalize the high frequency spin dynamics, resilience to external magnetic fields and lack of stray fields of antiferromagnetic materials for information storage and transfer[1,2]. The emerging novel subfield of antiferromagnetic magnonics seeks to make use of insulating antiferromagnets to transport angular momentum for information processing via magnons, the quanta of magnetic excitation[3] where magnons can be excited and transported via both the Néel vector (**n**) and potential net magnetic moments (**m**)[4]. In easy-axis antiferromagnets, spin angular momentum can be transferred by the circularly polarized magnon eigenmodes, enabling for instance long-distance transport of magnons in the low temperature easy-axis phase of the antiferromagnetic iron oxide, hematite ($\alpha$-$Fe_2O_3$)[5,6]. Easy-plane antiferromagnets on the other hand have linearly polarized magnon eigenmodes, which conventionally do not carry angular momentum[7,8]. However, a superposition of modes that dephase has been shown to transport magnons efficiently in the room temperature, easy-plane phase of $\alpha$-$Fe_2O_3$[9-11]. Although long-distance magnon transport has been shown in both easy-axis and easy-plane $\alpha$-$Fe_2O_3$, magnon transport exceeding decay-lengths of a few nanometers has yet to be demonstrated in any other antiferromagnet of either easy-axis or easy-plane anisotropy. This raises the question of whether there is something unique about hematite amongst the compendium of antiferromagnetic insulators or whether other antiferromagnets can exhibit long distance angular momentum transport and in particular what magnetic properties are required.

Although $\alpha$-$Fe_2O_3$ has shown to exhibit micrometer magnon decay lengths across a large temperature range, the transport efficiency drops rapidly across the easy-axis to easy-plane phase transition, the Morin transition[9]. Furthermore, the antiferromagnetic anisotropy in hematite is extremely temperature sensitive[12], meaning that propagating magnons experience regions of differing anisotropy in the presence of thermal gradients, limiting the potential for active manipulation of the magnon current by, for example, localized Joule heating[13]. It has been demonstrated that dilute doped hematite with a non-magnetic ion still facilitates efficient magnon transport despite the changes in magnetic symmetry that occur[14]. If this doping is instead taken to the extreme, where fifty percent of the Fe is substituted, a large class of materials, known as orthoferrites, appear and these are expressed with the formula $XFeO_3$. These materials have a range of interesting properties that can be explored in conjunction with magnon transport such as piezo-electricity[15-17] and a large magnetostriction[18,19]. For the most part, this class of materials are antiferromagnetic with a range of symmetries depending on the exact element $X$[20]. Due to Dzyaloshinskii-Moriya interaction (DMI), orthoferrites also typically have a net magnetization that arises due to canting of the magnetic sublattices, that appears perpendicular to the DMI vector, if the DMI is not parallel to the Néel vector[21,22]. Furthermore, this large class of materials can also exhibit potentially low damping, one of the key requirements for long distance spin transport. Thus, this materials class offers an exciting playground for probing the transport mechanisms of antiferromagnetic materials through a range of symmetries and anisotropies to unravel the critical criteria for efficient magnon propagation and control.

Here we demonstrate magnon transport over micrometers in the ultra-low damping, orthoferrite $YFeO_3$ making use of a novel transport mode unique to this type of system. This key observation of a new transport mode opens up a large number of technologically relevant materials for low power, antiferromagnetic magnonic devices. Using an external magnetic field to manipulate the magnon eigenmode polarization, we characterize the transport efficiency and identify the dominant role of the Néel vector. Although the zero-field magnon eigenmodes are non-degenerate, unlike in easy-axis hematite, they are linearly polarized and unable to facilitate zero-field magnon transport. On applying an external field, the magnon eigenmodes become elliptically polarized when the field has a nonzero projection on the Néel vector, thus enabling efficient magnonic spin current transport over large distances. Both the exponential decay of spin transport signal with distance and its vanishing magnitude at lower temperature indicate a diffusive nature to the transport. The transport is furthermore anisotropic and we demonstrate that the magnon decay length varies between the different crystallographic axes, which can be explained by the anisotropic magnon group velocity arising from anisotropic exchange stiffness.

Antiferromagnetic YFeO$_3$ ($T_N$ = 644 K)[20,23] crystallizes in an orthorhombic structure[24] and has three principal anisotropy axes. The [100] *a*-axis and [010] *b*-axis are the antiferromagnetic easy and hard axes respectively, whilst the [001] *c*-axis is an intermediate anisotropy axis (see Fig. 1(a)). A strong DMI of 14 T parallel to the *b*-axis leads to a net magnetic moment orientated along the *c*-axis[25,26]. An external magnetic field applied along the *a*-axis initiates a smooth rotation of **n** confined within the *ac*-plane until **n** aligns parallel to the *c*-axis at the critical field ($H_{cr}$). The value of $H_{cr}$ varies between 7.4 T and 6.5 T at temperatures between 4.2 and 300 K[27,28]. Such a rotation under an applied field is caused by the homogeneous DMI[21], and contrasts with the abrupt spin-flop transition typical in purely collinear antiferromagnets. The DMI found in YFeO$_3$ is significantly stronger than that found in bulk hematite[29] and correspondingly the net moment is larger (see supplementary). This also enables the possibility to disentangle the feasibility for antiferromagnetic transport dictated by the net magnetic moment. Unlike hematite, this material has no temperature-driven spin transition[30] and the antiferromagnetic anisotropy is approximately constant over a broad temperature range making it potentially more apt for devices. It has also been shown to have extremely fast domain wall motion[31-33] that offers the potential for magnon driven domain wall motion for information storage.

## Results

**Magnetic Damping in YFeO$_3$.** Before exploring non-local spin transport experiments, we first determine if the magnetic damping coefficient ($\alpha$), a key parameter for determining the spin-transport length scale, is low enough. To acquire information about the magnetization dynamics and $\alpha$, we characterize the magnetic resonance of a 0.5 mm thick [010]-oriented single crystal of YFeO$_3$ using a continuous-wave electron paramagnetic resonance spectrometer operating at high frequencies from 127 GHz to 510 GHz[9,34]. The antiferromagnetic resonance measurements show multiple peaks associated with magnetostatic modes (shown in supplementary). We extract the average linewidth by taking the sum of the individual linewidths divided by the number of peaks. Fig. 1(b) shows the frequency dependence of the average linewidth for the low frequency magnon eigenmode for the field directed along the *c*-axis. The linewidth dependence is fitted using the theoretical model of antiferromagnetic resonance provided by Fink[35], which yields an estimate for $\alpha$ of $9.3\pm2\times10^{-5}$. Hence, the magnetic damping of YFeO$_3$ is of the same order of YIG[36], the ferromagnetic material with lowest reported $\alpha$, and hematite[9], the only antiferromagnetic material which has previously shown long distance spin transport.

**Devices and detection technique.** To investigate the spin transport in [010]-oriented YFeO$_3$ (see methods and supplementary for growth and characterization details), we make use of devices (see Fig. 1(c)) with two different geometries; where Pt wires are parallel and perpendicular to the easy-axis. When a charge current (*I*) is driven through the injector Pt wire, a transverse spin current is generated due to the spin Hall effect (SHE), which yields a spin accumulation **μ**$_s$ at the Pt/YFeO$_3$ interface, polarized perpendicular to the flow direction of the charge current[37,38]. The spin current is absorbed by the antiferromagnet when **μ**$_s$ is parallel to **n**, resulting in a nonequilibrium distribution of magnons with an average spin parallel to **μ**$_s$. These magnons then diffuse in the YFeO$_3$, away from the injector. The magnon current flowing in the YFeO$_3$ is absorbed by a spatially and electrically separated Pt detector, where it is converted to a measurable voltage $V_{el}$ by the inverse-SHE[5,39]. As $V_{el}$ reverses its sign on reversing the current polarity, owing to the SHE[40] that reverses the spin current direction, we calculate the $V_{el}$ signal as ($V_{el}$ (I+) - $V_{el}$ (I-))/2 to remove polarity-independent effects[5,41]. We then express the spin transport signal as the nonlocal resistance $R_{el} = V_{el}/I$ (in units of Ohm).

**Spin transport for wires along the easy-axis.** We first consider the spin transport in devices where the Pt-wires are parallel to the easy-axis (Fig. 2(a)). This geometry places the interfacial spin polarization **μ<sub>s</sub>** perpendicular to **n** in the absence of a magnetic field (**H**). On sweeping the field along the easy-axis, **n** smoothly rotates due to the DMI within the *ac*-plane[21,22], reaching the *c*-axis (which is perpendicular to field) at $H_{cr}$ (= 6.5 T) where it remains for $H > H_{cr}$. From the previous reports on hematite we expect that with the increased projection of **n** on **μ<sub>s</sub>** with increasing field along the easy-axis, the $R_{el}$ signal should correspondingly increase and reach a maximum at $H_{cr}$[5,6]. As we increase the magnetic field applied to the YFeO$_3$ crystal, we see a gradual increase of $R_{el}$ as **n** rotates, however, contrary to hematite we observe a maximum below $H_{cr}$, followed by a sharp, sudden decrease before the signal vanishes at $H_{cr}$ and remains constant around zero within the error bar (shown in Fig. 2(a)). Therefore, the spin transport signal appears in the intermediate field value between **n** parallel and perpendicular to field. If instead, the field is applied perpendicular to the easy-axis, parallel to the **m** along the *c*-axis, no $R_{el}$ signal is observed across the whole field range (shown in supplementary), which can be understood from the fact that **n** always remains perpendicular to both field and transport direction unlike in hematite. Hence, the response of the spin-transport signal by applying field along and perpendicular to easy-axis indicates that transport of spin information is mediated only by the Néel order. Any transport mediated by the net magnetic moment is negligible and remains within the experimental error. Thus, the transport is purely antiferromagnetic in nature. The solid line in Fig. 2(a) is the fit using theoretical modelling discussed later and in the supplementary, considering the spin angular momentum only mediated by the Néel vector, showing excellent agreement.

**Spin transport for wires perpendicular to the easy-axis.** Having shown that the Néel vector facilitates efficient magnon transport along the intermediate anisotropy *c*-axis under an applied magnetic field, we next turn to investigating the spin transport along a different crystallographic axis, i.e., *a*-axis (the easy-axis) in order to probe the role of the anisotropy. To do this, we choose device where wires are perpendicular to the easy-axis. In such a configuration, at zero field, **n** is parallel to **μ$_s$**. Previous reports on hematite show zero-field spin transport in such a configuration, as the magnon eigenmodes are circularly polarized at zero field and **n** has a finite projection on **μ$_s$**[5,6]. However, in stark contrast, we find that in YFeO$_3$ there is no significant zero-field spin transport. On increasing the field along the easy-axis, the $R_{el}$ signal increases and reaches a maximum at $H = 3$ T followed by a decrease and the signal vanishes at $H_{cr}$ (shown in Fig. 2(b)). The mechanism for the increase in $R_{el}$ to a maximum and the decrease with increasing field can be explored by the equilibrium orientation of the **n**, as described in previous geometry (shown in Fig. 2(a)). In this device geometry, we observe that the peak appears at lower fields than for the previous geometry as the spin-transport signal depends on two factors; the field dependence of the magnon magnetization and the projection of **n** on **μ$_s$**. This behaviour is in line with the theoretical expectations as described in the section on the theoretical model below. Furthermore, on applying the field perpendicular to the easy-axis, no spin-transport signal is observed (shown in the supplementary), which indicates that the magnetic field perpendicular to **n** is unable to generate the required ellipticity to the magnon eigenmodes to enable magnon transport.

We also probed the field dependence of thermal magnon transport for both device geometries and establish that the transport of thermally-excited magnons is also mediated dominantly by Néel vector below $H_{cr}$, which is the sharp contrast to hematite where thermal spin transport is mainly mediated by field induced magnetic moment[5] (details are discussed in supplementary).

Thus, as a first key result we can conclude that the angular momentum transport in YFeO$_3$ is significantly different from the previously observed transport in hematite with major qualitative differences. In particular, the dominance of the spin transport on the Néel vector orientation and magnon ellipticity is further confirmed by additional measurements varying the field direction with respect to the anisotropy axes (see supplementary for details).

**Estimation of magnon decay length.** To check if the transport mode present in YFeO$_3$ also lends itself to long distance transport, we measure the magnon decay length ($\lambda$) that reflects the efficiency of the spin transport. With the exception of hematite, all reported antiferromagnets have so far yielded $\lambda < 10$ nm[42-45]. To determine $\lambda$, we have measured $R_{el}$ as a function of centre-to-centre distance between the wires up to 1.25 μm for both of the geometries as shown in Fig. 3. $R_{el}$ is recorded at the maximum signal for each geometry i.e., at 5.5 T for devices shown in Fig. 2a and at 3 T for devices shown in Fig. 2b. Furthermore, to compare $\lambda$ along two anisotropy axes at a constant field, in order to rule out a possible field dependent reduction of $\lambda$ such as observed in ferromagnetic materials[46], $R_{el}$ is also recorded at 3 T for the former geometry. The curves are fitted with an exponential decay function: $R_{el} = Ae^{-d/\lambda}$, where $A$ is a distance independent pre-factor, and yield $\lambda = 470 \pm 40$ nm and $525 \pm 50$ nm at 5.5 T and 3 T, respectively, for the former device and $280 \pm 20$ nm at 3 T for the latter device. The exponential fitting indicates a diffusive nature of the transport[5,6,39] and excludes the possibility of dominating spin superfluidity[47-49]. Also, the temperature dependence of $R_{el}$ (shown in the supplementary) at low temperature points to the diffusive nature of the transport[9].

We observe a significant difference in magnon decay length scale along different anisotropy axes, where $\lambda$ is ~ 50% lower along the *a*-axis than the *c*-axis. For ferromagnets, such an effect has been explained by anisotropy in the magnetic energy under the influence of magnetic dipole-dipole interaction[50]. In the case of antiferromagnets and weak ferromagnets dipole-dipole interactions are negligible. Hence, the anisotropy in the magnetic energy does not affect the group velocity of magnons, as is the case in ferro- or ferrimagnets. The observed anisotropy of $\lambda$ in our case can be explained by the anisotropy of the magnetic damping (or relaxation time) or anisotropy of the magnon group velocity along different anisotropy axes. Using the theoretical analysis discussed in detail below, we identify the anisotropy in the magnon group velocity as the main mechanism for the observed anisotropic magnon decay length (see inset in Fig. 3).

## Discussion

**Theoretical model of magnon spin transport in YFeO$_3$.** To understand and interpret the magnon transport in orthoferrites, we first calculate the magnon eigenmodes by solving standard equations for antiferromagnetic dynamics in the presence of an external magnetic field (details in supplementary). We then analyze the angular momenta of each mode, associating them with the dynamic magnetization $\mathbf{m}_{dyn} = M_s \delta \mathbf{n} \times \delta \dot{\mathbf{n}} / \gamma H_{ex}$.

The magnon spectra include two branches with frequencies,

$$\omega_{1,2} = \sqrt{\left(\omega_{1,2}^0(H)\right)^2 + c_a^2 k_x^2 + c_b^2 k_y^2 + c_c^2 k_z^2}, \tag{1}$$

where the $\omega_{1,2}^0(H)$ are field-dependent frequencies of antiferromagnetic resonance (see Fig. 4(a)), $c_a \neq c_b \neq c_c$ are limiting magnon velocities along different crystallographic directions (*a*, *b*, and *c*), and **k** is the magnon wave-vector. $\omega_{1,2}^0(H)$ depends on the material parameters, described in the supplementary. In the absence of an external field, all the eigenmodes are linearly polarized and correspond to oscillations of the Néel vector along either the *c*-axis (low-frequency eigenmodes $\omega_1(\mathbf{k})$) or along the *b*-axis (high-frequency eigenmodes $\omega_2(\mathbf{k})$). The magnetization of such modes oscillates with the magnon frequency and cannot transport time-independent (dc) angular momentum, yielding a zero-transport signal at zero field, as experimentally observed (and in contrast to hematite where zero-field transport is present). The linear polarization of the eigenmodes is a direct consequence of the strong orthorhombicity in which case $\omega_1^0 \neq \omega_2^0$ due to the different magnetic anisotropies along the *a*- and *b*-axes. It should be noted that the magnon velocity is also anisotropic and depends on the direction of **k**. However, this is related with anisotropy of the exchange stiffness (different exchange coupling along different axes[51], where $J_c > J_{ab}$, so, we can assume that $c_c > c_a$).

The external magnetic field $\mathbf{H} \| \boldsymbol{a}$ applied along the easy-axis induces a smooth reorientation of the Néel vector, confined to the *ac*-plane. Figure 4(b) shows the projection of the **n** along the *a*-axis as a function of the field along *a*-axis, where the projection of *H* on the *b*-axis is zero. In this geometry, the magnetic field not only affects the values of $\omega_{1,2}^0(H)$, it also modifies the polarization of the eigenmodes from linearly polarized to elliptically polarized. The magnetization of the elliptically polarized modes also has a dc component, which is parallel to equilibrium orientation of the Néel vector, $\mathbf{m}_{dyn} \| \mathbf{n}^{(0)}$, and whose value is proportional to the product of magnon density, magnon frequency and magnon ellipticity (see Fig. 4(c)). All three contributions depend on the magnetic field. Consequently, an increase of the spin-transport signal with field is observed. The direction of the magnetization in the low-frequency magnon branch corresponds to $\mathbf{m}_{dyn} \cdot \mathbf{H} > 0$ and is opposite for the high-frequency branch. At the critical field $H_{cr}$, corresponding to the alignment of the Néel vector along the *c*-axis (perpendicular to the magnetic field), the ellipticity of the modes goes to zero and above $H \geq H_{cr}$ magnons are again linearly polarized and unable to transfer dc angular momentum.

The electrical, $R_{el}$, (and thermal, $R_{th}$, in the supplementary) transport signals are proportional to the weighted sum of the dc magnetizations of all available magnon eigenmodes. In addition, $R_{el} \propto (\mathbf{n}^{(0)} \cdot \hat{\mu})^2$ which corresponds to the projection of the $\mathbf{m}_{dyn}$ on the direction of the spin accumulation $\hat{\mu} || \boldsymbol{\mu}_s$ ($|\hat{\mu}| = 1$) during the pumping and detection of non-equilibrium magnons[9]. $R_{th} \propto \mathbf{n}^{(0)} \cdot \hat{\mu}$, as the magnetization is projected only during the detection. Assuming that contribution of different magnon modes is proportional to the thermal probability given by Bose distribution, we calculate the expected $R_{el}$ signal which we show as solid line fits to the experimental data in Fig. 2.

Next, we consider the diffusion of magnons that is responsible for the transport of angular momentum. The spin diffusion coefficient $D$ is calculated in the framework of linear nonequilibrium thermodynamics as a response to the gradient of spin accumulation $\boldsymbol{\mu}_s$ (see Ref. 4). The corresponding expression reads as:

$$D = \frac{1}{T} \int \frac{d^3 k}{(2\pi)^3} \sum_{\alpha=1,2} \tau_\alpha(\mathbf{k}) \left( v_{x\alpha} \cos\psi + v_{y\alpha} \sin\psi \right)^2 \frac{(\mathbf{m}_\alpha \hat{\mu}) \exp(\hbar \omega_\alpha / T)}{(\exp(\hbar \omega_\alpha / T) - 1)^2} \quad (2)$$

where $\tau_\alpha(\mathbf{k})$ and $\mathbf{v}_\alpha \equiv \partial \omega_\alpha / \partial \mathbf{k}$ are the relaxation time and group velocity of the magnon of the $\alpha$-mode with the wave vector $\mathbf{k}$, respectively. The angle $\psi$, calculated from the $a$-axis, defines the direction of the spin accumulation gradient and $T$ is the temperature. Using the dispersions given in equation (1) and assuming that the relaxation time is the same for all magnon eigenmodes, we determine that the value of the diffusion coefficient $D$ is anisotropic, $D \propto c_a^2 \cos^2\psi + c_c^2 \sin^2\psi$. The inset of Fig. 3 shows the angular dependence of $\lambda \propto \sqrt{D} \propto \sqrt{c_a^2 \cos^2\psi + c_c^2 \sin^2\psi}$ calculated based on the assumption that $c_c \approx 38$ km/sec $> c_a \approx 20$ km/sec, which is based on experimental observations[52]. From this, we conclude that the anisotropy observed in the magnon-decay length originates mainly from anisotropy of the exchange stiffness (or, at the microscopic level, from anisotropy of the exchange interactions in the $ab$-plane and along the $c$-axis[51]), though the anisotropy of the relaxation time can also contribute to the observed effect. This mechanism contrasts with the mechanism of the spin diffusion anisotropy in ferro- and ferrimagnets, which originates mainly from the long-range dipole-dipole interaction[50]. In antiferromagnets and in weak ferromagnets like hematite and orthoferrites, the net magnetization is small and the dipole-dipole interactions can thus be neglected.

The observed long-distance spin transport in the ultra-low damping, insulating canted antiferromagnet $YFeO_3$ is an important step forward towards the ultimate goal of establishing a universal model for long distance spin transport in antiferromagnetic insulators. Our findings open up a large and important class of low damping materials in which one can propagate spin information. The transport mechanism previously identified in hematite is not universal to all antiferromagnets and entail limitations: in easy-axis antiferromagnets, only intrinsic circularly polarized eigenmodes can carry spin angular momentum and in easy-plane antiferromagnets, a superposition of two linearly polarized eigenmodes can support spin transport, but the transport length scales are dominated by the dephasing length. The reported new mechanism in $YFeO_3$ allows us to transport spin by modifying the ellipticity of magnon modes. Hence, we emphasize the fact that all the antiferromagnets with a low magnetic damping have the potential to transport spin if one tunes the ellipticity of the magnon eigenmodes. For this, there are different possible approaches and here we demonstrate that we can use a field to modify the ellipticity in the presence of strong DMI. Finally, we point out that whether the presence of strong DMI is essential to efficiently modify the magnon ellipticity with an applied field is an open question that warrants study in further materials. Furthermore, the observed new mode of transport demonstrated by nonlocal transport measurements over long distances along with the antiferromagnetic resonance measurements highlights the potential of low damping antiferromagnetic insulators for their integration into next generation magnonic and spintronic devices.

## Methods

**Single crystal preparation.** The YFeO$_3$[010] crystal is prepared by optical floating zone technique from the sintered polycrystalline sample. The crystallographic properties have been investigated by x-ray diffraction, which confirm the crystallinity of the crystal and also the angle of miscut between the crystallographic axes and sample plane. The details of the growth and characterization are discussed in the supplementary. The magnetization hysteresis curve is measured using superconducting quantum interference device to detect the weak magnetic moment along the intermediate anisotropy *c*-axis.

**Non-local transport measurements.** Prior to the pattering, the sample was cleaned with acetone, isopropanol and deionized water to remove any organic surface residue. The devices were patterned using e-beam lithography followed by sputter deposition of 7-nm Pt and lift-off. The electrical contact pads were defined by a second e-beam lithography step followed by the deposition of Cr (5 nm)/Au (45 nm) and lift-off. Devices are consisting of three wires of length 50 μm and width 300 nm. The center-to-center separation between the wires varies from 500 nm to 2.5 μm. A scanning electron microscope (SEM) image of a typical device is shown in Fig. 2(c). The sample was mounted to a piezo-rotating element in a variable temperature insert installed in a superconducting magnet, which is capable of fields up to 12 Tesla (T). We pass a charge current through the central wire with a charge density of $3.8 \times 10^7$ A/cm$^2$ and measure a nonlocal voltage at the detector wires. The nonlocal voltage is recorded for positive and negative polarity of current as a function of field, spatial distance between the wires and the angle between current and field directions.

## Data availability

The data that support the findings of this study are available from the corresponding authors upon reasonable request. Correspondence and material request should be addressed to M.K. or S.X.C.

## Acknowledgements

S.D. thanks Mr. T. Reimer of Johannes Gutenberg University Mainz for his help in fabricating the devices. This work was supported by the Max Planck Graduate Center with the Johannes Gutenberg-Universität Mainz (MPGC). The authors in Mainz acknowledge support from the DFG project number 423441604. R.L. and M.K. acknowledge financial support from the Horizon 2020 Framework Programme of the European Commission under FET-Open grant agreement no. 863155 (s-Nebula). All authors from Mainz also acknowledge support from both MaHoJeRo (DAAD Spintronics network, project number 57334897 and 57524834), SPIN + X (DFG SFB TRR 173 No. 268565370, projects A01, A03, A11, B02, and B12), and KAUST (OSR-2019-CRG8-4048.2). M.K. acknowledges support by the Research Council of Norway through its Centers of Excellence funding scheme, project number 262633 "QuSpin". S.B acknowledges the Deutsche Forschungsgemeinschaft (DFG, German Research Foundation) - project number 358671374. S.X.C. acknowledges support by the Science and Technology Commission of Shanghai Municipality (No.21JC1402600), and the National Natural Science Foundation of China (NSFC, No. 12074242).

## Authors Contribution

M.K. and A.R. conceived the idea. X.X.M. and H.Y.C. grew, cut, and characterized the YFeO$_3$ single crystals used in the experiments under the guidance of S.X.C. S.B. performed the structural and magnetic characterization. S.D. fabricated the devices and M.A.S. deposited Pt. The non-local transport measurements are performed by S.D. with the help of S.B., C.S. and F.F. Magnetic resonance measurements are performed by F.V.D., U.E., V.B., A.L.B. and R.L. The theoretical model is developed by O.G. The data are analyzed by S.D., M.K., A.R., R.L., S.B., O.G., J.S. and G. J. The manuscript is written by S.D. with input from A.R. O.G. R.L and M.K. All the authors revised the manuscript.

## Competing interests

The authors declare no competing interests.

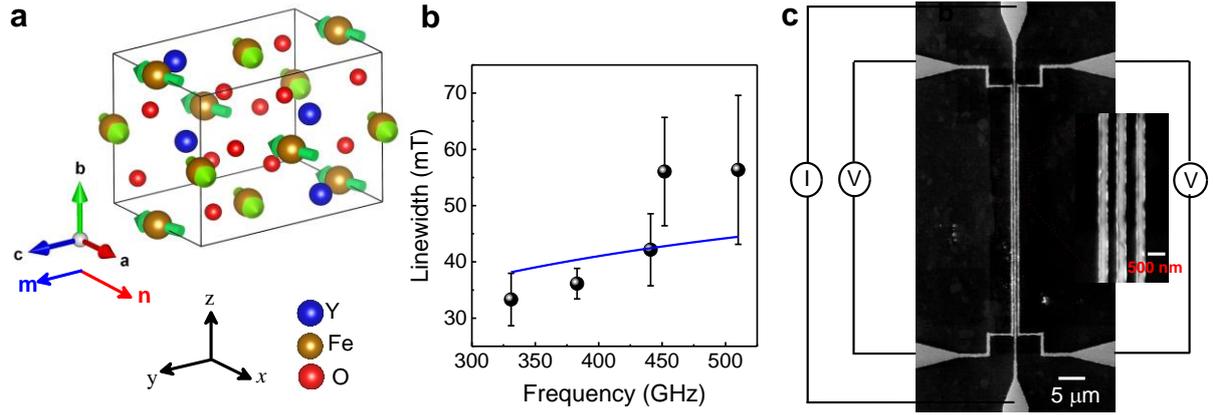

Fig. 1: **Crystal structure, magnetic resonance and the device.** (**a**) Crystal structures of YFeO$_3$[010], where the nearest neighbour Fe-ions are coupled antiferromagnetically along the easy-axis ([100] *a*-axis) and the moments are slightly canted along the *c*-axis which yield a net weak magnetization directed parallel to the *c*-axis. The arrows in the structure indicate the direction of spins of the Fe ions. (**b**) The average resonance linewidth as a function of frequency for the low frequency magnon mode for a field applied along the *c*-axis of a [010]-oriented YFeO$_3$ single crystal. The blue line is the theoretical fitting using the model from Fink.[35], which yields $\alpha$ of $9.3 \pm 2 \times 10^{-5}$. (**c**) SEM image of a typical device, where the charge current is driven along the middle wire and the non-local voltages are measured in both wires to the left and right of it. We associate orthogonal coordinates with the crystallographic axes: **x** ∥ **a**, **y** ∥ **c**, **z** ∥ **b**.

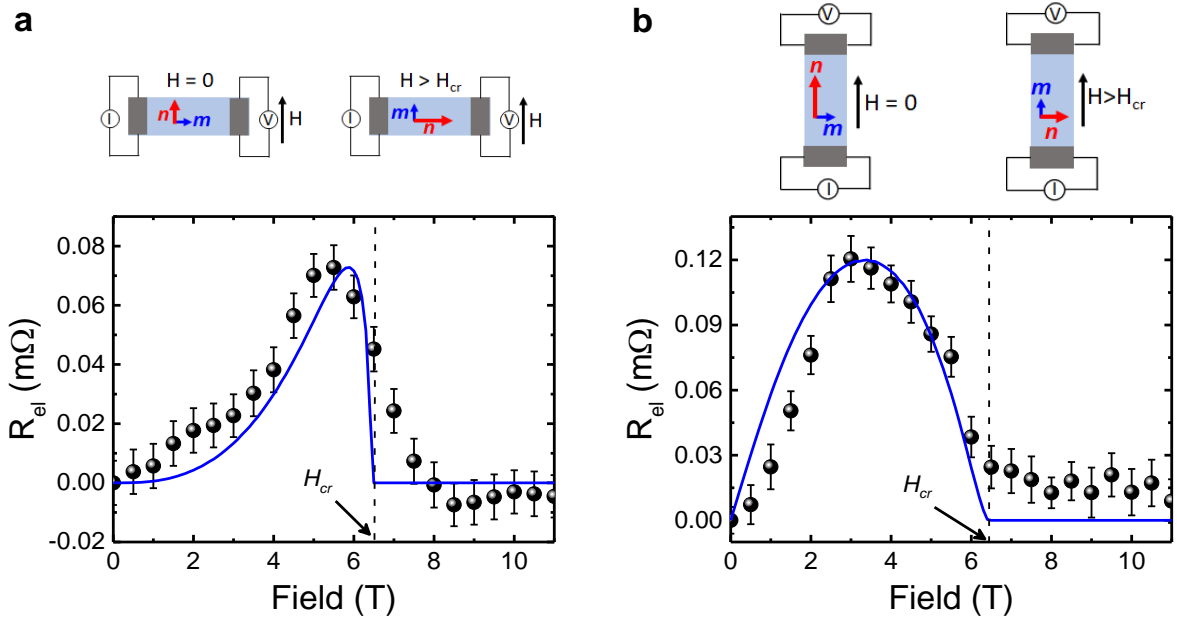

Fig. 2: **Spin transport in YFeO$_3$ at 200 K.** (**a**) $R_{el}$ as a function of field applied along easy-axis for device where wires are parallel to easy-axis. Initially **n** is parallel to the field and perpendicular to **μ**$_s$, which shows zero spin transport signal. On increasing the field **n** rotates smoothly yielding a gradual increase in $R_{el}$, followed by a maximum and the signal vanishes above $H_{cr}$ where **n** becomes perpendicular to the field. (**b**) $R_{el}$ as a function of field applied along the easy-axis for the device where the wires are perpendicular to easy-axis. In spite of **n** parallel to **μ**$_s$ at zero magnetic field, no zero-field spin transport is observed. $R_{el}$ increases with field initially and reaches a maximum at $H = 3$ T followed by a decrease and the signal diminishes above critical field. The error bars are the standard error of the mean. In both curves, the signal is plotted after subtracting the offset at zero field. The solid blue lines based on the theoretical model are described in the text. In both of the geometries, the centre-to-centre distance between the wires is about 525 nm.

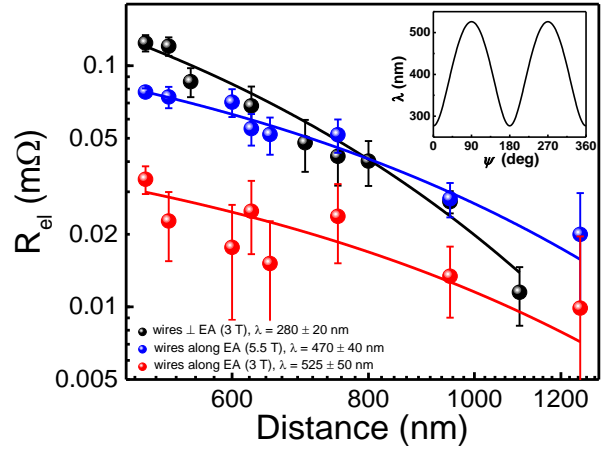

Fig. 3: **Distance dependence of spin transport signal at 200 K.** Nonlocal spin transport signal $R_{el}$ as a function of centre-to- centre distance between of the wires for devices where wires are along the easy-axis at $H$ = 5.5 T and 3 T and for devices where wires are perpendicular to the easy-axis at $H$ = 3 T. The error bars are standard errors of the mean. The solid lines are the fit to an exponential decay function which yields value of $\lambda$. Inset, theoretical calculation of $\lambda$ as a function of angle $\psi$ measured from the easy-axis, considering the experimental value of $\lambda$ along and perpendicular to easy-axis.

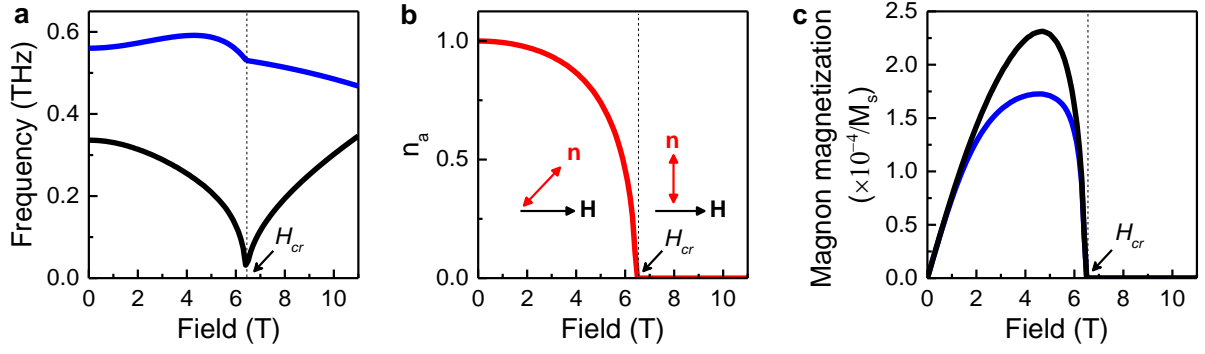

Fig. 4: **Characteristics of two magnon eigenmodes.** (**a**) Field dependence of the eigen-frequencies for two magnon eigenmodes. Softening of the low frequency eigenmode takes place when the magnetic field is parallel to the *a*-axis at $H = H_{cr}$. (**b**) Component of Néel vector along *a*-axis as a function of field applied along easy-axis, indicating a smooth orientation of Néel vector towards the *c*-axis. The critical field is defined at a field where $n_a$ is zero. (**c**) Integral contribution of all magnons into magnetization averaged over the **k**-space with the Bose distribution function for low-frequency and high-frequency branches. For fields along the easy-axis, the ellipticity is zero above $H_{cr}$, where the $\mathbf{n} \perp \mathbf{H}$ and the magnon modes are linearly polarized.

# Supplementary - Anisotropic long-range spin transport in canted antiferromagnetic orthoferrite YFeO$_3$

Shubhankar Das[1], A. Ross[2,$], X. X. Ma[3,$], S. Becker[1], C. Schmitt[1], F. van Duijn[4,5], F. Fuhrmann[1], M.-A. Syskaki[1], U. Ebels[4], V. Baltz[4], A.-L. Barra[5], H. Y. Chen[3], G. Jakob[1,6], S. X. Cao[3,§], J. Sinova[1], O. Gomonay[1], R. Lebrun[2], and M. Kläui[1,6,7*]

[1]Institute of Physics, Johannes Gutenberg University Mainz, Staudingerweg 7, 55128 Mainz, Germany

[2]Unité Mixte de Physique CNRS, Thales, Université Paris-Saclay, Palaiseau 91767, France

[3]Department of Physics, Materials Genome Institute, International Center for Quantum and Molecular Structures, Shanghai University, Shanghai, 200444, China

[4]Univ. Grenoble Alpes, CNRS, CEA, Grenoble INP, SPINTEC, F-38000 Grenoble, France

[5]Laboratoire National des Champs Magnétiques Intenses, CNRS-UGA-UPS-INSA-EMFL, F-38042 Grenoble, France

[6]Graduate School of Excellence Materials Science in Mainz, Staudingerweg 9, 55128 Mainz, Germany

[7]Center for Quantum Spintronics, Norwegian University of Science and Technology, Trondheim 7491, Norway

*klaeui@uni-mainz.de

§sxcao@shu.edu.cn

$these authors contributed equally

**Single crystal sample preparation.** $Y_2O_3$ (4N) and $Fe_2O_3$ (3N) powders were used as raw materials to make the polycrystalline sample according to the stoichiometric ratio by the conventional solid-state reaction method. The original reagents were weighed carefully and pulverized with moderate anhydrous ethanol in an agate mortar. The mixtures were sintered at 1250 °C for 1000 minutes to guarantee an adequate reaction and then furnace-cooled to room temperature. To ensure sufficient reaction, we continue to grind the pre-sintered sample into powders, and then it is pressed into a thin sheet of about 1.5 mm thickness to narrow the gap between the powder particles for secondary sintering. The sintering temperature and duration are the same as for the pre-fired process. The secondary sintered samples were thoroughly reground, and the polycrystalline powders were pressed into two rods that are 35-85 mm in length and 5-6 mm in diameter by Hydrostatic Press System at 60 MPa, and then sintered again at 1250 °C for sufficient reaction.

High-quality $YFeO_3$ single crystals were then successfully grown by an optical floating zone furnace (Crystal System Crop, model FZ-T-10000-H-VI-P-SH) from the sintered polycrystalline sample. In the process of crystal growth, both the upper and lower rods rotate at the opposite direction of 15 rpm at the same time, and the molten zone moves upward at a speed of 2 mm/h. The picture of the $YFeO_3$ single crystal is shown in Fig. S1.

The crystallographic orientations of the $YFeO_3$ single crystal were determined by back-reflection Laue X-ray photography. Clear and sharp Laue spots imply a high quality of the single crystal. The back-reflection Laue images of the $YFeO_3$ single crystal along the *a*-axis, *b*-axis, and *c*-axis are shown in Fig. S2. Then samples of $YFeO_3$ single crystal perpendicular to the different symmetry axes are cut from the single crystal, whose accurate direction is further confirmed by four-cycle X-ray diffraction (XRD) measurement.

**Structural characterisation.** The crystallographic properties were investigated using XRD. The sample has been fixed to an aluminium plate and measured in a Bruker D8 equipped with a copper anode, Göbel mirror and a monochromator. The sample has first been aligned with respect to the sample surface in the setup. The (010) reflex has then been found by tilting the sample by 11° out of the initial position. This indicates that the sample has a miscut. As a result, the c- and a-axes are also misaligned with the surface by around 8° as shown in Fig. S3 (a). The crystallographic purity has been confirmed by performing 2θ/ω measurements along the (010), (110) and (021) direction. An example scan along the (010) direction is shown in Fig. S3 (b). No other peaks except the expected $YFeO_3$ peaks are found in these scans, indicating the absence of foreign phases or twinned crystallographic domains. The calculated lattice parameters of $a = 5.28$ Å, $b = 5.57$ Å and $c = 7.6$ Å are in line with other reports[1].

**Magnetic Characterization.** To check the magnetic properties of the $YFeO_3$ crystal, *m-H* hysteresis curves have been measured at 300 K by applying the field along the c-axis (see Fig. S4). We observe a saturation magnetization of around 0.05 $\mu_B$/f.u., which is relatively large compared to that of bulk $\alpha$-$Fe_2O_3$, that can be understood from the large Dzyaloshinskii–Moriya interaction field of 16 T leading to a canting of the antiferromagnetically aligned sublattices. A sharp transition between the magnetization direction is observed with a coercive field of 45 ± 5 mT.

**Antiferromagnetic resonance.** Figure S5(a) shows the resonance peak at around 7 T field of the low frequency magnon eigenmode measured at a frequency of 452 GHz. The multiple resonance peaks appear because the wavelength is far smaller than the thickness of the crystal. We extract the average linewidth by taking the sum of the individual linewidth divided by the number of peaks. Figure S5(b) shows frequency vs. resonance field for the low frequency magnon mode. The frequency dependence is fitted using the model presented in Ref. 2.

**Thermal spin transport for field applied along the easy-axis.** The flowing charge current in the Pt-injector leads to Joule heating, creating a lateral temperature gradient between the Pt wires. This gradient results in a thermal spin current induced by the spin Seebeck effect[3,4]. The thermal magnon spin current flowing in the YFeO$_3$ is then absorbed by a spatially and electrically separated Pt detector, where it is converted to a measurable voltage $V_{th}$ by the inverse-SHE[5,6]. As $V_{th}$ does not depend on current polarity in the injector, we calculate the $V_{th}$ signal by adding the voltage at the detector due to the opposite polarity current in injector[6,7]; $V_{th} = (V(I+) + V(I-))/2$. We then express the non-local signal due to thermal transport as $R_{th} = V_{th}/I^2$ (in units of VA$^{-2}$).

Figure S6(a) shows thermal spin signal $R_{th}$ as a function of field for device where wires are along the easy-axis. On applying the magnetic field, $R_{th}$ increases and reaches maximum at a field strength below $H_{cr}$, followed by a sharp decrease on further increasing the field. The peak and the decrease in $R_{th}$ replicate the electrical transport signal $R_{el}$ (see Fig. 2(a) of main text) and can be explained from the equilibrium orientation of Néel vector (**n**) as described in the main text. This reveals that the dominant contribution of thermal spin transport is also mediated by Néel vector. Both the non-zero $R_{th}$ signal and the linear increase with field above $H_{cr}$ are not directly related to the spin orientation but could have origins in emerging field induced magnetic moments, Hanle effect, thermal Hall effect, Seebeck effect and interfacial spin Seebeck effect at the detector etc.[8-11]. The necessary data to be able to disentangle every single effect goes beyond the scope of this work and could be a unique study alone.

Figure S6(b) shows field dependence of $R_{th}$ for device where wires are perpendicular to the easy-axis. The $R_{th}$ shows distinct behaviour; starting with zero $R_{th}$ signal in absence of field, it increases significantly with field followed by a maximum at $H = 3$ T, then turn to decrease and shows a change of sign. Finally, at $H > H_{cr}$ the gradient of the signal reverses and signal amplitude linearly increases with field. The change in sign of $R_{th}$ at 6 T and the linear increase above $H_{cr}$ are the outcome of the emergent dominant contribution from various mechanisms mentioned in previous geometry (shown in Fig. S6(a)). The contribution to $R_{th}$ from those various mechanisms is opposite in sign to the contribution coming from Néel vector in this geometry[8-11]. The solid lines in Fig. S6(a) and (b) indicate the transport expectation from our model, showing excellent agreement.

**Spin transport for field perpendicular to the easy-axis.** Figure S7 (a) and (b) show the $R_{el}$ and $R_{th}$ signal as a function of field applied perpendicular to the easy-axis for the device where wires are parallel to the easy-axis. No $R_{el}$ signal is observed in the whole field range, whereas the gradual increase of $R_{th}$ and saturation tendency at high field indicates that the field induced net moment contributes to thermal spin transport[8-10].

Figure S8 (a) and (b) show $R_{el}$ and $R_{th}$ signals as a function of field applied perpendicular to the easy-axis for the device where wires are perpendicular to the easy-axis. On applying the field perpendicular to the easy-axis, we didn't observe any $R_{el}$ signal due to the non-rotation of **n** which prefers to stay perpendicular to the field. This indicates that the field perpendicular to **n** is unable to create the ellipticity in the magnon modes that is required to transport spin angular momentum (see discussion section in the main text and below for details). A very weak linear field dependence of $R_{th}$ is observed which may arise from the misalignment of crystallographic axes with the sample plane due to crystal miscut (see Fig. S3(a)).

**Angular dependence of spin transport for wires perpendicular to the easy-axis.** Figure S9(a) and (b) show the angular dependence of the $R_{el}$ and $R_{th}$ signal, respectively, for various fields, where $\alpha$ is defined as the angle between the current direction ($c$-axis) and the field. The $R_{th}$ vs $\alpha$ curve shows the expected $\cos(\alpha)$ dependence at 3 T, but the angular dependence below and above the critical field show a reverse gradient. The solid lines are the fitting based on magnon dynamics considered in the proposed theoretical model.

**Temperature dependence of the spin transport.** We have studied the field dependence of spin transport signal for various temperatures for device where wires perpendicular to the easy-axis. Figure S10 shows the $R_{el}$ signal as a function of temperature at $H = 3$ T (the field where maximum of $R_{el}$ vs. $H$ curve is observed). With increasing the temperature, $R_{el}$ signal increases and shows a maximum value around 200 K and further decreases with higher temperature. The peak around 200 K can be explained from the temperature dependence competition between magnon decay length and magnon population. With decreasing temperature, the magnon decay length increases[12] whereas the magnon population decreases. We could not detect any signal below 50 K within our measurement accuracy. The absence of detectable signal at low temperature indicates the diffusive nature of the transport and excludes spin-superfluidity as the dominant mode of transport.

## Theoretical Model

For the description of the magnetic dynamics, we use the same approach as suggested in Ref. 12. We use the standard dynamical equation for the Néel vector (**n**, |**n**| = **1**), which is valid on the assumption of small spin canting

$$\mathbf{n} \times \left[ \ddot{\mathbf{n}} - 2\gamma \dot{\mathbf{n}} \times \mathbf{H} + \gamma \alpha_G H_{ex} \dot{\mathbf{n}} - \nabla \cdot \hat{c}^2 \nabla \mathbf{n} + \gamma^2 \frac{H_{ex}}{M_s} \frac{\partial w_{AF}}{\partial \mathbf{n}} \right] = \gamma^2 H_{ex} \mathbf{n} \times \mathbf{H}_{curr} \times \mathbf{n}. \qquad (1)$$

Here $\hat{c} = c_a, c_b, c_c$ is the diagonal tensor of the limiting magnon velocity along different crystallographic axes, $\gamma$ is the gyromagnetic ratio, $\alpha_G$ is the Gilbert damping coefficient, $H_{ex}$ is the exchange field that keeps the magnetic sublattice moments antiparallel, $\mathbf{H}_{curr} = \hbar \varepsilon \theta_H \mathbf{j} \times \hat{z} / (2 e d_{AF} M_s)$ is directed along the spin accumulation direction $\hat{\mu}$ ($|\hat{\mu}| = 1$) of the current in the Pt electrode, **j** is the current density, $\hbar$ is the Planck constant, $d_{AF}$ is the penetration depth of the spin current into the orthoferrite, $0 < \varepsilon \le 1$ is the spin-polarization efficiency, $\theta_H$ is the spin Hall angle, $e$ is the electron charge, and $M_s/2$ is sublattice magnetization. The expression for the magnetic energy density $w_{AF}$ (**n**; **H**) in the presence of the constant external magnetic field **H** can be written as[13,14]

$$w_{\mathrm{AF}}(\mathbf{n}) = -\frac{1}{2}M_s H_a n_x^2 + \frac{1}{2}M_s H_b n_z^2 + \frac{M_s}{2H_{ex}}(\mathbf{H}\cdot\mathbf{n})^2 + \frac{M_s}{H_{ex}}H_{DMI}\hat{z}\cdot\mathbf{n}\times\mathbf{H}, \tag{2}$$

where $H_a > 0$ and $H_b > 0$ are anisotropy fields associated with the easy (*a*-axis) and hard (*b*-axis) magnetic direction. $H_{DMI} > 0$ is the homogeneous DMI field responsible for a small spin canting (and finite magnetization) along the *c*-axis. Orthogonal coordinates are associated with the crystallographic axes: $\hat{x} \parallel \mathbf{a}, \hat{y} \parallel \mathbf{c}, \hat{z} \parallel \mathbf{b}$.

For modelling we use the following values of the parameters: $H_{DMI}$ = 16 T, $H_a$ = 0.225 T, $H_{cr}$ = 6.5 T, and $H_{ex}$ = 640 T, consistent with the reported values[13]. The anisotropy field along the *b*-axis was estimated as $H_b$ = 0.625 T, to satisfy the ratio between low and high frequencies magnon eigenmodes observed in Ref. 15,16.

**Equilibrium state.** The equilibrium orientation of the Néel vector, $\mathbf{n}^{(0)}(\mathbf{H})$, in the presence of the field applied parallel to the easy-axis, $\mathbf{H} \parallel \hat{x}$, rotates smoothly within the *xy* plane[13], so that

$$n_y^{(0)}(H_x) = \frac{H_{DMI} H_x}{H_a H_{ex} - H_x^2}. \tag{3}$$

At critical field

$$H_{cr} = -\frac{1}{2}H_{DMI} + \sqrt{H_a H_{ex} + \frac{1}{4}H_{DMI}^2}, \tag{4}$$

the Néel vector reaches perpendicular to the magnetic field and parallel to the intermediate anisotropy axis $c \parallel \hat{y}$. Further increasing the magnetic field ($H_x > H_{cr}$) produces only a canting of magnetic sublattices resulting in a nonzero magnetization,

$$m_x = M_s \frac{H_{DMI} + H_x}{H_{ex}}. \tag{5}$$

This behaviour of the Néel vector contrasts with the abrupt spin-flop transition, which is observed in compensated antiferromagnets, and is governed by DMI.

**Spin transport.** To study the magnon spin transport, we start from calculations of magnon spectra in presence of the external magnetic field. For this we consider small fluctuations $\delta \mathbf{n}$ of the Néel vector on top of an equilibrium state $\mathbf{n}^{(0)}$: $\mathbf{n} = \mathbf{n}^{(0)}(\mathbf{H}) + \delta \mathbf{n}$, $\delta \mathbf{n} \perp \mathbf{n}^{(0)}$ and assume that $\delta \mathbf{n}(t, \mathbf{k}) \propto \exp(-i\omega t + i\mathbf{k} \cdot \mathbf{r})$. The magnons are then calculated as eigenmodes of the linearized equation (1) that takes the form

$$\delta \ddot{n}_1 - 2\omega_H \delta \dot{n}_2 - \sum_{j=1}^{3} c_j^2 \frac{\partial^2}{\partial x_j^2} \delta n_1 + \omega_1^2 \delta n_1 = 0$$

$$\delta \ddot{n}_2 + 2\omega_H \delta \dot{n}_1 - \sum_{j=1}^{3} c_j^2 \frac{\partial^2}{\partial x_j^2} \delta n_2 + \omega_2^2 \delta n_2 = 0, \tag{6}$$

where $\delta n_{1,2}$ and $\omega_{1,2}^2$ are the eigen-vectors and eigen-values of the matrix $\gamma^2 H_{ex} M_s \left( \partial^2 w_{AF} / \partial n_j \partial n_k \right)\big|_{\mathbf{n}^{(0)}}$, $\omega_H = \gamma \mathbf{H} \cdot \mathbf{n}^{(0)}$. The eigen-frequencies of two magnons eigenmodes with $\mathbf{k} = 0$ are given by the expression

$$\omega_\pm^2 = \frac{1}{2}\left(\omega_1^2 + \omega_2^2\right) + 2\omega_H^2 \pm \sqrt{\frac{1}{4}\left(\omega_1^2 - \omega_2^2\right)^2 + 2\omega_H^2\left(\omega_1^2 + \omega_2^2\right) + 4\omega_H^4}. \tag{7}$$

Frequencies of the modes with nonzero $\mathbf{k}$ are obtained by substitution $\omega_{1,2}^2 \to \omega_{1,2}^2 + \sum_{j=1}^{3} c_j^2 k_j^2$. This approach is appropriate for magnons with $\mathbf{k}$ vectors far from the Brillouin zone edge which give the main contribution to the observed spin transport. Figure 4(a) of the main text shows the field dependence of eigen-frequencies $\omega_\pm(H)$ calculated for field along the easy-axis consistent with the previous reports[15]. The low frequency mode vanishes at the critical field $H = H_{cr}$, and the high frequency mode has nonzero frequency which slightly decreases above $H_{cr}$.

Next, we discuss the ability of eigen modes to transport spin. The angular momentum of an eigenmode is a vector parallel to the dynamical magnetization[17]

$$\mathbf{m}_{dyn} = M_s \frac{\delta \mathbf{n} \times \delta \dot{\mathbf{n}}}{\gamma H_{ex}}. \tag{8}$$

From the orthogonality condition $\delta \mathbf{n} \perp \mathbf{n}^{(0)}$, it follows that $\mathbf{m}_{dyn} \parallel \mathbf{n}^{(0)}$. We consider only the modes with stationary magnetization, as only these modes contribute to spin transport signal.

As follows from equation (8), magnon magnetization depends on the polarization of the magnon mode. We introduce the polarization $0 \leq |\varepsilon_{\pm}| \leq 1$ as the ellipticity of the mode, as will be specified below (see equation (9)). The maximal polarization corresponds to circularly polarized modes with an ellipticity $\varepsilon_{\pm} = 1$. The minimal polarization corresponds to linearly polarized modes with an ellipticity $\varepsilon_{\pm} = 0$. Calculations based on equation (6) show that the eigenmodes in presence of the field are circularly-polarized with field-dependent ellipticity

$$\varepsilon_{\pm} = \frac{4\omega_H \omega_{\pm} \left( \omega_{\pm}^2 - \omega_1^2 \right)}{\left( \omega_{\pm}^2 - \omega_1^2 \right)^2 + 4\omega_H^2 \omega_{\pm}^2}. \tag{9}$$

Each of the modes can carry field dependent angular momentum proportional to magnetization $m_{\pm} = \omega_{\pm} \varepsilon_{\pm} M_s / \gamma H_{ex}$. In contrast to the uniaxial case, ellipticities of the modes are different, $|\varepsilon_+| \neq |\varepsilon_-|$. The magnetizations of the modes differ not only in sign, but also in value, due to the difference of the frequencies. It should be noted that the absolute value of the angular momentum is defined by the intensity of magnon fluctuations ($\propto \delta \mathbf{n}^2$) and depends on the external parameters (temperature or spin current).

**Nonlocal transport.** We associate $R_{el}$ contribution with the current induced magnon spin current and $R_{th}$ with spin Seebeck effect. The spin current in the Pt electrode modifies the effective damping coefficient of the spin-polarized modes and creates a nonequilibrium distribution of magnons with average magnetization $\mathbf{m}_{ave}$[12]

$$\mathbf{m}_{ave} = \frac{\gamma}{\alpha_G} M_s \mathbf{n}^{(0)} \left( \mathbf{H}_{curr} \cdot \mathbf{n}^{(0)} \right) \left[ \varepsilon_+ \omega_+ f\left(\frac{\hbar \omega_+}{T}\right) + \varepsilon_- \omega_- f\left(\frac{\hbar \omega_-}{T}\right) \right], \quad (10)$$

where $f(\varepsilon) = \left[ \exp(\varepsilon/(k_B T)) - 1 \right]^{-1}$ is the Bose-Einstein equilibrium distribution function for each of the modes, $\hbar$ is the Planck constant, $k_B$ is the Boltzmann constant. The gradient of the spin accumulation induces diffusion of spin polarized magnons to the detector electrode and is detected by the inverse spin-Hall effect. In this case $R_{el} \propto \mathbf{m}_{ave} \cdot \mathbf{j}_{ISHE} \times \hat{z} \propto \left(\mathbf{n}^{(0)} \cdot \hat{\mu}\right)^2$, where $\mathbf{j}_{ISHE}$ is the current density in the detector electrode. As both electrodes for spin-pumping and measuring are parallel, $\mathbf{j}_{ISHE} \times \hat{z} \parallel \mathbf{H}_{curr} \parallel \hat{\mu}$.

Spin Seebeck effect is induced by the temperature gradient between the electrodes due to the Joule heating. In this case magnetization of the magnon is related with the field induced ellipticity of the magnon eigenmodes:

$$\mathbf{m}_{ave} \propto \frac{\gamma}{\alpha_G} \mathbf{n}^{(0)} \left[ \varepsilon_+ \omega_+ f\left(\frac{\hbar \omega_+}{T}\right) - \varepsilon_- \omega_- f\left(\frac{\hbar \omega_-}{T}\right) \right]. \quad (11)$$

Correspondingly, $R_{th} \propto \mathbf{m}_{ave} \cdot \mathbf{j}_{ISHE} \times \hat{z} \propto \left(\mathbf{n}^{(0)} \cdot \hat{\mu}\right)$.

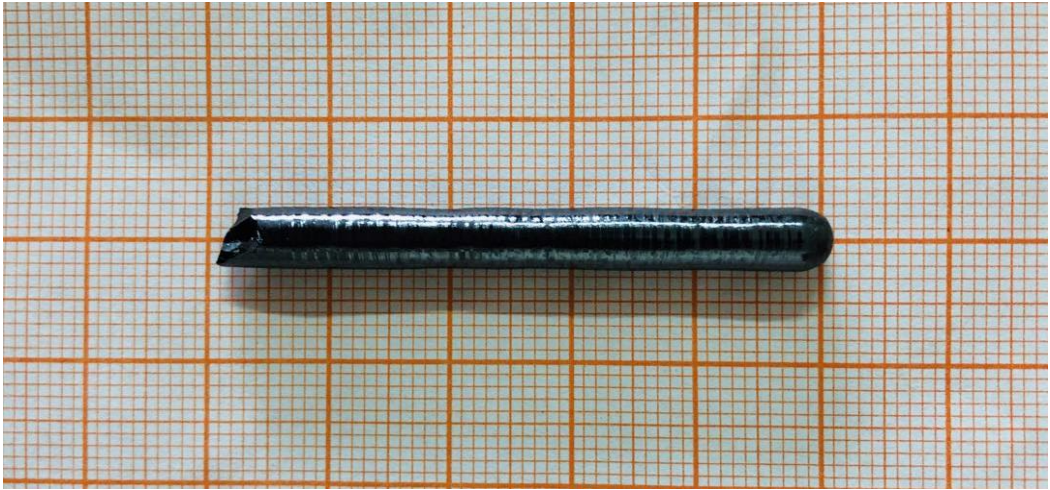

Fig. S1: **Single Crystal.** Picture of YFeO$_3$ single crystal grown by optical float-zone method.

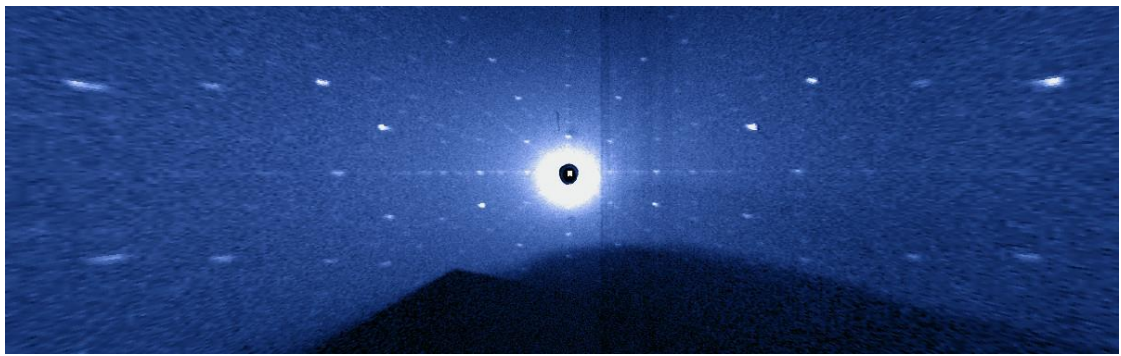

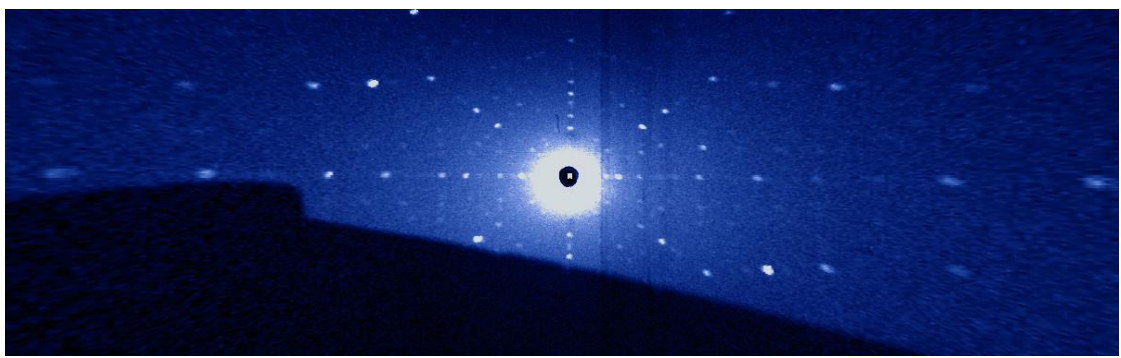

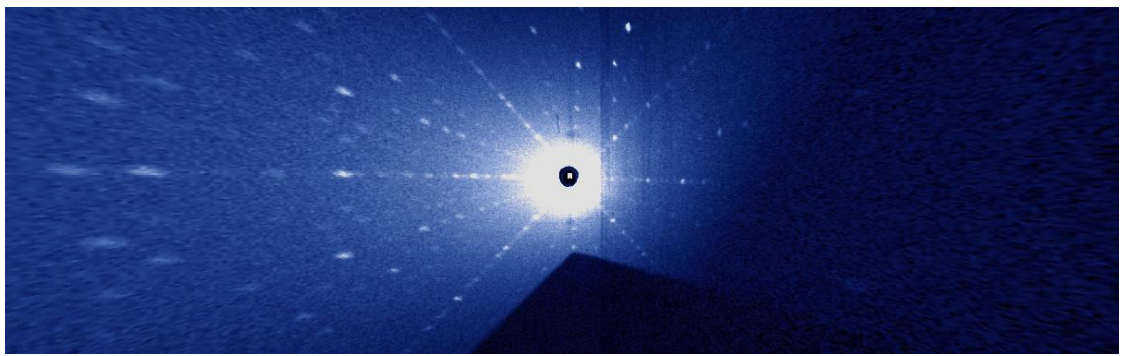

Fig. S2: **Laue X-ray photography.** Laue photography of YFeO$_3$ single crystal along the (**a**) *a*-, (**b**) *b*-, and (**c**) *c*- axis.

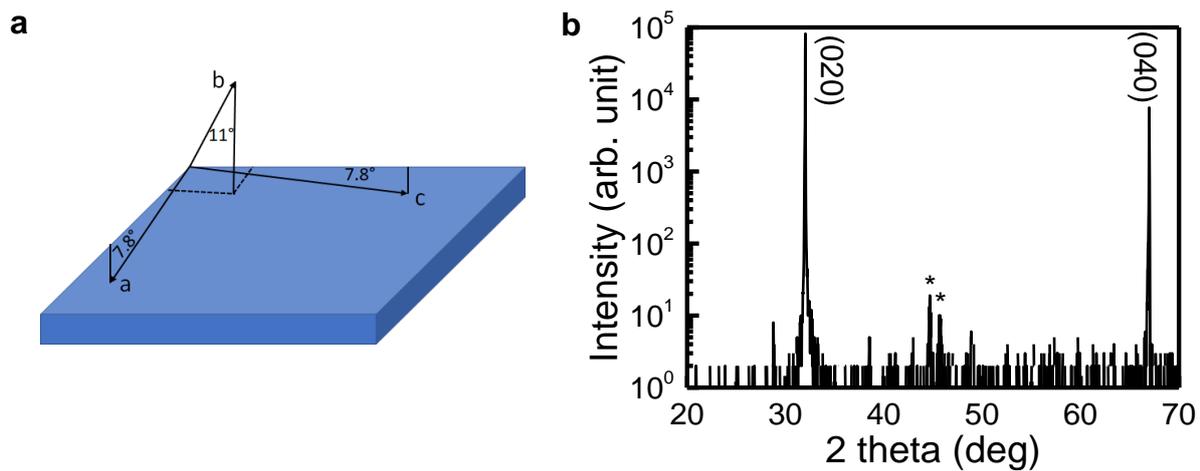

Fig. S3: **Sample orientation.** (**a**) Schematics of crystallographic axes with respect to sample plane due to miscut. (**b**) $\theta$-$2\theta$ scan measured along (010) direction perpendicular to the surface of the sample plane. Only (020) and (040) peaks are observed and levelled accordingly. Peaks marked with a star come from aluminium sample holder and are not related to the YFeO$_3$ crystal.

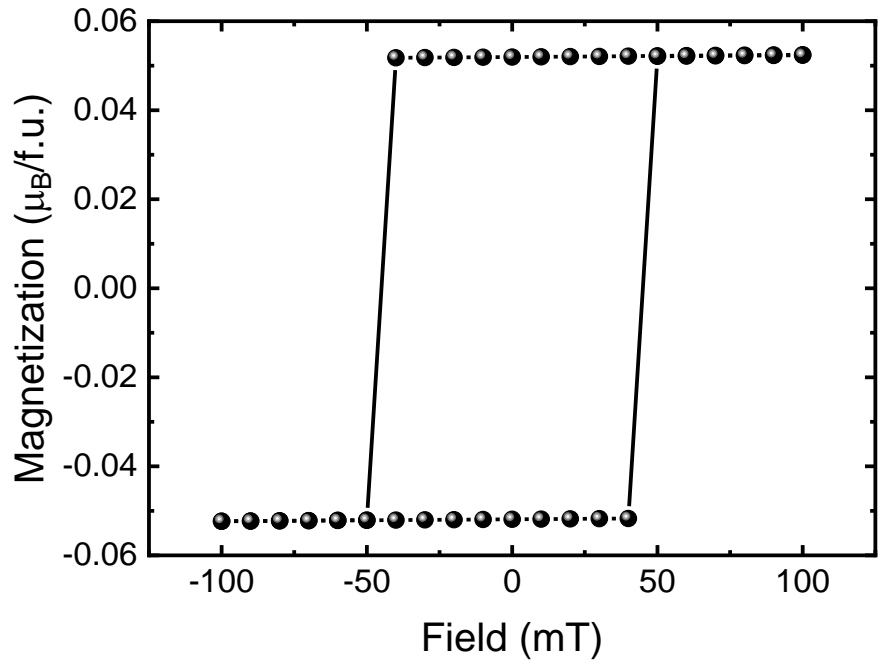

Fig. S4: **Magnetization curve.** The *m-H* hysteresis loop for field applied along *c*-axis.

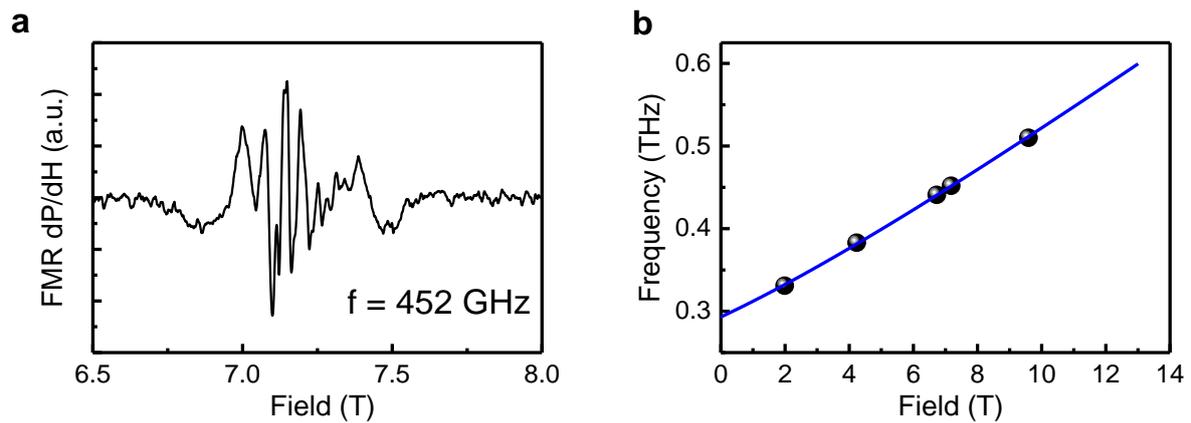

Fig. S5: **Magnetic resonance.** (**a**) Resonance peak as a function of magnetic field at 452 GHz for the 0.5 mm thick YFeO$_3$ single crystal. (**b**) Resonance frequency as a function of field for the low frequency magnon mode.

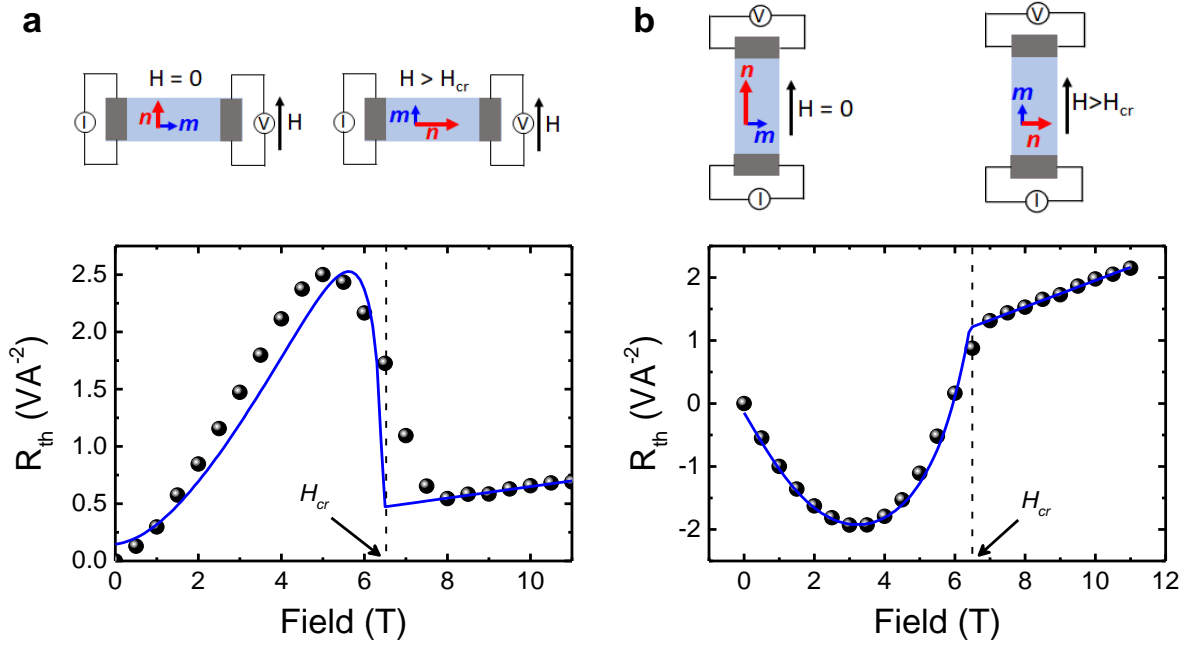

Fig. S6: **Thermal spin transport at 200 K.** (**a**) $R_{th}$ as function of field applied along the easy-axis in device where wires are parallel to the easy-axis. $R_{th}$ increases with field and reaches maximum below $H_{cr}$ followed by a sharp decrease and above $H_{cr}$ a linear field dependence is observed. (**b**) $R_{th}$ as function of field applied along the easy-axis in device where wires are perpendicular to the easy-axis. $R_{th}$ increase with field to reach a maximum at 3 T followed by a change of sign below $H_{cr}$ and finally a change in gradient above $H_{cr}$. In both curves, the signal is plotted after subtracting the offset at zero field value. The solid lines correspond to the theoretical model are described in the text. The error bars are smaller than the symbols.

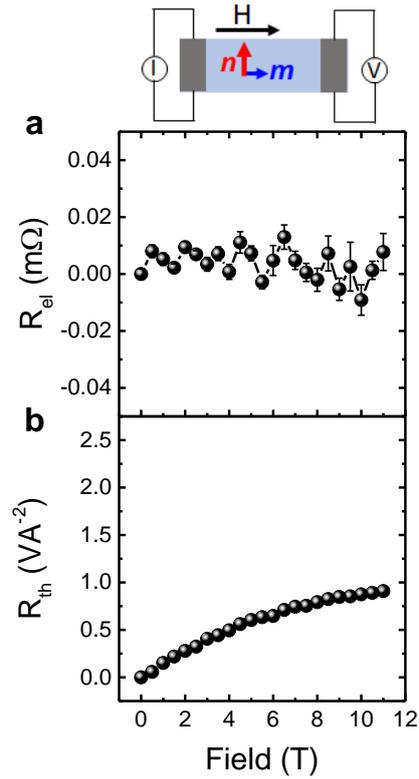

Fig. S7: **Spin transport for wires along the easy-axis and field perpendicular to the easy-axis.** (a) $R_{el}$ signal is plotted as a function of field perpendicular to the easy-axis. No spin transport signal is observed in the whole field range. The error bars are calculated from the standard deviation of the mean. (b) $R_{th}$ as a function of field shows a significant increase with field and a saturation tendency at higher field. The error bars are within the symbol height. The schematics is showing the direction of Néel vector, magnetization and field direction with respect to the wire.

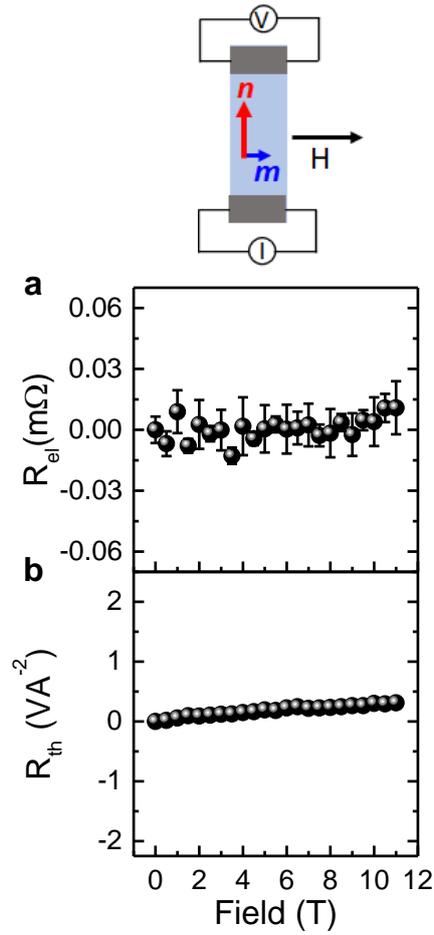

Fig. S8: **Spin transport for wires perpendicular to the easy-axis and field perpendicular to the easy-axis.** Schematics shows the relative orientation of the wire, Néel vector, magnetization and applied magnetic field. (**a**) $R_{el}$ vs H, where no spin signal is observed in whole field range from 0 to 11 T. (b) Weak increase in $R_{th}$ as a function of field.

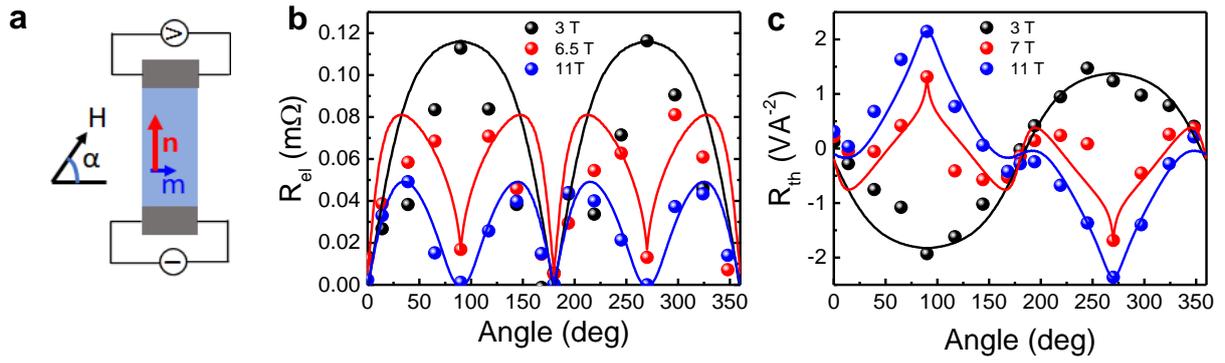

Fig. S9: **Angular dependence of spin transport signal.** (**a**) Schematics of the device and the measurement geometry along with the direction of Néel vector and weak magnetization at zero magnetic field are shown. (**b**) and (**c**) Angular dependence of $R_{el}$ and $R_{th}$ signal, respectively, for various magnetic fields. The solid lines are the fitting based on magnon dynamics from the proposed theoretical model. A constant offset is subtracted from the experimental data.

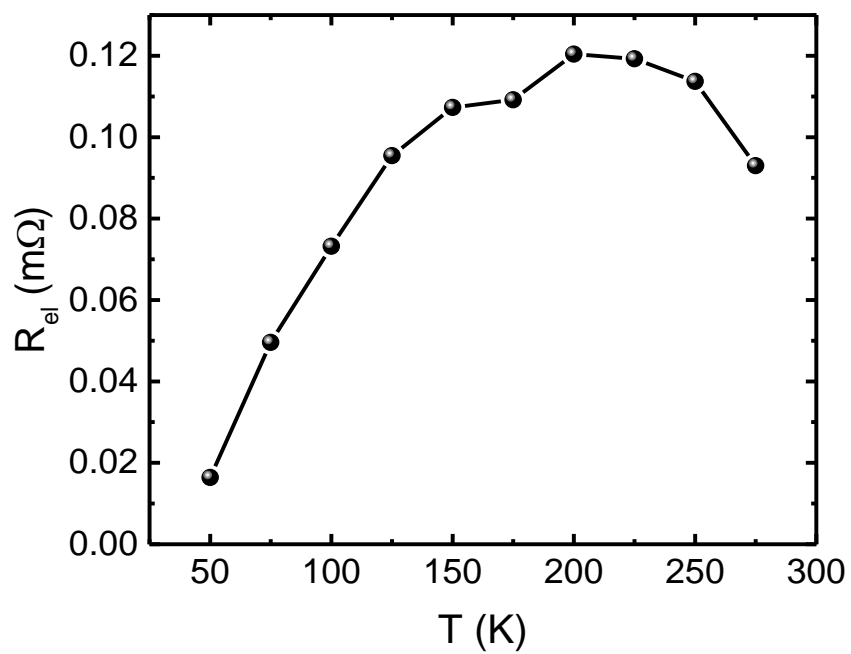

Fig. S10: **Temperature dependence of spin transport.** Spin transport signal as a function of temperature at $H = 3$ T for device where wires are perpendicular to the easy-axis.